\documentclass[aps,prb,twocolumn]{revtex4}
\usepackage{graphicx,bm}
\begin{document}
\title{Dynamics of a domain wall in a magnetic nanostrip: a toy model}
\author{D. Clarke}
\author{G.-W. Chern}
\author{O. A. Tretiakov}
\author{O. Tchernyshyov}
\affiliation{Johns Hopkins University, Department of Physics and
Astronomy, 3400 N. Charles St., Baltimore, Maryland 21218}

\begin{abstract}
In this report we demonstrate a simple model for the motion of a
vortex domain wall in a ferromagnetic strip of submicron width under
the influence of an external magnetic field.  The model exhibits three
distinct dynamical regimes.  In a viscous regime at low fields the
wall moves rigidly with a velocity proportional to the field.  Above a
critical field the motion becomes underdamped as the vortex
moves periodically across the strip; these oscillations are accompanied
by a slow drift with a decreasing velocity.  At still higher fields
the drift velocity starts rising linearly with the field again but
with a much lower mobility $dv/dH$ than in the low-field regime.
We calculate the relevant quantities and compare them to experimentally
observed values.
\end{abstract}

\maketitle

\section{Introduction}

Dynamics of domain walls in ferromagnetic strips and rings with submicron
dimensions is a subject of active research.\cite{Allwood05,Thomas07,Chien07}
This topic is directly relevant to several proposed schemes of magnetic
memory and is also interesting from the standpoint of basic physics.
The dynamics of domain walls under an applied magnetic field has distinct
regimes: viscous motion with a relatively high mobility at low fields
and underdamped oscillations with a slow drift at higher fields.\cite{Beach05}

The nontrivial dynamics is related to the composite nature of a domain wall
in a nanostrip: it consists of a few---typically two or three---elementary
topological defects in the bulk and at the edge of the strip.  As a result,
a domain wall has several low-energy degrees of freedom that are relevant
to the dynamics.  Weak external perturbations engage only the softest
(zero) mode---rigid translations along the strip.  Larger external forces
excite higher modes thereby altering the character of motion.

The general approach to the dynamics of domain walls in thin ferromagnetic
strips with a submicron width $w$ and thickness $t \ll w$ was described
recently by Tretiakov \textit{et al.}\cite{Tretiakov07a}
The configuration of a domain wall is parametrized by a few collective
coordinates
$\bm \xi = \{\xi_1, \xi_2, \ldots, \xi_N\}$ and the free energy of the
system $U$ is treated as a function of $\bm \xi$.  The resulting equations
of motion can be written in the vector notation as
\begin{equation}
\mathbf F - \hat \Gamma \dot{\bm \xi} + \hat G \dot{\bm \xi} = 0.
\label{eq-main}
\end{equation}
Here components of the vector $\mathbf F$ are generalized forces
$F_i = -\partial U/\partial \xi_i$; the symmetric matrix $\hat \Gamma$
and antisymmetric matrix $\hat G$ represent the viscous and gyrotropic
tensors, respectively.

The main goal of this paper is to illustrate the
collective-coordinate approach on a very simple model of a vortex
domain wall\cite{Chern-unpub} that served as a prototype for a
more realistic model of Youk \textit{et al}.\cite{Youk05}
Despite its simplicity, the model captures all of the main features
of a vortex domain wall and yields simple analytical results for the
relevant physical quantities. Quantitatively speaking, the values of
the forces computed in this model deviate by no more than 30\% from
those obtained in the more realistic model of the vortex wall.  Thus
one can make meaningful comparisons between the analytical results
obtained in this paper and experimental data.

In the main body of the paper we describe the simplified model of the wall
and compute the generalized forces and the viscous and gyrotropic tensors.
By substituting these quantities into Eq.~(\ref{eq-main}) we obtain simple
equations of motion.  At low fields the equations describe steady viscous
motion of the wall with a velocity proportional to the applied field.
The vortex is shifted in the transverse direction by
an amount proportional to the velocity of the wall.  At a critical velocity
the vortex is expelled from the strip and the steady motion breaks down
giving way to an oscillatory regime.  As the applied field increases further,
the drift velocity decreases at first but then again becomes proportional
to the applied field; the mobility coefficient $dv/dH$ is substantially
lower than the corresponding value in the viscous regime at low fields.
These results are compared to experimental data.

\section{model wall}\label{model}

\begin{figure}
\includegraphics[width=0.9\columnwidth]{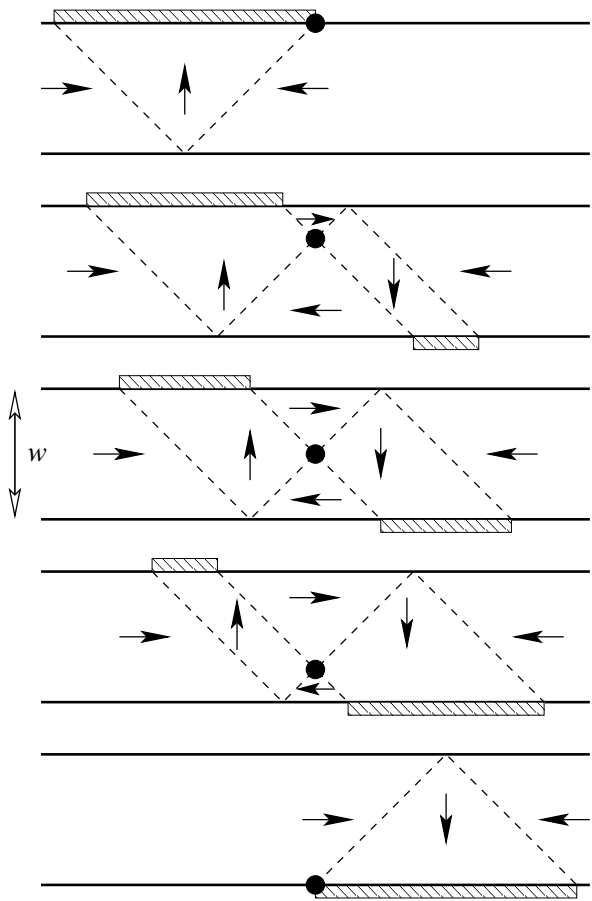}
\caption{A simple model for a vortex domain wall. \cite{Chern-unpub} The
panels show states with a fixed longitudinal coordinate of the vortex
$X = \mathrm{const}$; the transverse coordinate is $Y = w/2$, $w/4$,
0, $-w/4$, and $-w/2$.  The vortex core is denoted by
the filled circle.  Shaded areas indicate the locations of magnetic
charges.} \label{fig-gw}
\end{figure}

In our calculations of the wall dynamics, we use a simple model of
the vortex domain wall consisting of four domains with uniform
magnetization and separated by $90^\circ$ Neel walls
(Fig.~\ref{fig-gw}).

We assume that only two softest modes of the vortex wall are
involved in magnetization dynamics, so that the configuration of the
wall is fully described by the two coordinates $(X,Y)$ of the vortex
core.  In that case the equations of motion (\ref{eq-main}) reflect
the balance of forces acting on a particle moving in two dimensions
with a velocity $\mathbf V = (\dot{X}, \dot{Y})$.  The forces
include a conservative term $(-\partial U/\partial X, -\partial
U/\partial Y)$, a viscous term $\hat{\Gamma} \mathbf V$, and a
gyrotropic term\cite{Thiele73} $\hat{G} \mathbf V = pG \hat{\mathbf
z} \times \mathbf V$.  The gyrotropic force depends on the
out-of-plane polarization of the vortex core
$p=M^\mathrm{core}_z/|M^\mathrm{core}_z|$ and the gyrotropic
constant $G = 2\pi J t$, where $J = \mu_0 M/\gamma$ is the density
of angular momentum.\cite{Tretiakov07a}  The conservative and
viscous terms are discussed next.

\subsection{Free energy $U$ and conservative forces}

In strips that support vortex domain walls, the dominant
contribution to the energy in the absence of an applied field is due
to magnetostatic interactions. For any domain wall in a strip of
width $w$, thickness $t$, and saturation magnetization $M$, there is
a total magnetic charge $Q=2\mu_0 M t w$ associated with the
wall.\cite{Youk05} In a vortex wall, nearly all of this charge is
expelled to the edge. In our simple model, magnetic charges form two
lines of lengths $w-2Y$ and $w+2Y$ with constant density of charge
per unit length $\rho = \mu_0 Mt$.

The magnetostatic energy of this wall $E(Y) = E(0) + kY^2/2 +
\mathcal O(Y^4)$ has a minimum at $Y=0$. This leads to a force $-kY$
that acts to keep the vortex centered on the strip. In general, the
total exchange energy of the wall may change with the position of
the vortex, altering the restoring force slightly. However, in our
simplified model, the exchange cost comes entirely from the four
Neel walls that make up the vortex wall. Because the length of these
walls does not change as the vortex moves, we need not consider the
exchange interaction in our analysis of the wall dynamics.

A line of charges of length $L$ has the self-energy
\begin{equation}
E_0(L) = \frac{\mu_0 M^2 t^2}{8\pi} \int_{0}^L \int_0^{L}  \frac{dx
\, dx'}{|x-x'|}.
\end{equation}
The divergence at $x=x'$ requires a regularization.  In a crude way
this can be done by introducing a short-range cutoff in the
integral, i.e. by integrating over distances $|x-x'|>Ct$, where
$C$ is a numerical constant.  We then obtain a
logarithmic dependence on $Ct$:
\begin{equation}
E_0(L) = \frac{\mu_0 M^2 t^2}{4\pi} L [ \log{(L/Ct)} -1].
\end{equation}
The self-energies of the two lines of charge is
\begin{eqnarray}
E_\mathrm{self}(Y) &=& E_0(w-2Y) + E_0(w+2Y) \nonumber\\
&=& E_\mathrm{self}(0) +\frac{\mu_0 M^2 t^2 Y^2}{\pi w} + \mathcal
O(Y^4). \label{eq-self}
\end{eqnarray}
Note that the cutoff $C t$ affects only the constant term
$E_\mathrm{self}(0)$; the quadratic term is not sensitive to the
exact value of $C$.

In a similar way we evaluate the interaction of the two lines of
charges,
\begin{eqnarray}
E_\mathrm{int}(Y) &=& \frac{\mu_0 M^2 t^2}{4\pi}
\int_{-3w/2}^{2Y-w/2} \int_{2Y+w/2}^{3w/2} \frac{dx \,
dx'}{\sqrt{w^2+(x-x')^2}} \nonumber\\ &=&E_\mathrm{int}(0) -
\frac{\mu_0 M^2 t^2 Y^2}{\pi w\sqrt{5}} + \mathcal O(Y^4).
\label{eq-int}
\end{eqnarray}
The sum of the quadratic terms in Eqs.~(\ref{eq-self}) and
(\ref{eq-int}) yields the ``spring'' energy $kY^2/2$, from which we
determine the spring constant:
\begin{equation}
k = \frac{2(1-1/\sqrt{5})}{\pi} \frac{\mu_0 M^2 t^2}{w}.
\label{eq-k}
\end{equation}

Next we deal with the Zeeman energy of the wall $-\mu_0 t\int d^2x
\,  \mathbf H \cdot \mathbf M$ in the presence of an applied
magnetic field $\mathbf H$ parallel to the axis of the strip. A
longitudinal shift of the vortex by $\Delta X$ results in a pure
translation of the wall.  Independently of the wall shape, the rigid
shift changes the Zeeman energy by $-QH \Delta X$, where
$Q=2\mu_0Mtw$ is the magnetic charge of the wall.  Therefore the
longitudinal Zeeman force is $QH$ in any model.

The Zeeman force also has a transverse component.  As can be seen from
Fig.~\ref{fig-gw}, transverse motion of the vortex core changes the total
magnetization $M_x$ of the strip and thus affects its Zeeman energy.
As the vortex core crosses the strip from top to bottom (Fig.~\ref{fig-gw}),
the Zeeman energy decreases linearly by $4\mu_0 HM t w^2$.  Therefore
the transverse component of the Zeeman force is $-2QH$.

The total free energy of a wall with the vortex core at $(X,Y)$
is thus
\begin{equation}
U(X,Y)=kY^2/2+2QHY-QHX.
\label{eq-U}
\end{equation}

\subsection{Viscosity tensor $\hat{\Gamma}$ and viscous drag}

We next consider the viscosity of the vortex wall. The viscosity
that appears in Eq.~(\ref{eq-main}) is a symmetric matrix whose
components are given by:\cite{Tretiakov07a}
\begin{equation}
    \Gamma_{ij}=\alpha J t \int d^2x \,
    \frac{\partial \phi}{\partial\xi_i}
    \frac{\partial \phi}{\partial\xi_j},
\label{eq-Gamma-def}
\end{equation}
where $\phi$ is the azimuthal angle characterizing magnetization.

An infinitesimal shift in the collective coordinates $X$ and $Y$
affects magnetization in the vicinity of the Neel walls only.
We begin by considering the contribution of a single Neel wall
emanating from the vortex core at $\pm 45^\circ$, $\phi(x,y,X,Y) =
f(x-X \mp y \pm Y)$.  For such a wall, derivatives with respect to
collective coordinates can be reduced to ordinary gradients:
$\partial \phi/\partial X = -\partial \phi/\partial x = -f'$ and
$\partial \phi/\partial Y = -\partial \phi/\partial y = \pm f'$.
As a result, the tensor components are equal to each other, up to
a sign:
\[\Gamma_{XX} = \Gamma_{YY} = \mp \Gamma_{XY} = \alpha J t\int d^2x \,
{f'}^2. \] Note that this represents, up to a trivial constant, the
exchange energy of the Neel wall, which has been calculated, e.g.,
in Ref. \onlinecite{Chern05}.  We thus obtain viscosity coefficients
for the Neel walls intersecting at the vortex core, $\Gamma_{XX} =
\Gamma_{YY} = \mp \Gamma_{XY} = 0.152 \alpha J t w/\lambda,$ where
the exchange length $\lambda = \sqrt{A/\mu_0 M^2} = 3.8$ nm in
permalloy.  Opposite signs of the off-diagonal component
$\Gamma_{XY}$ can be understood by noting that, as the vortex moves
along $Y$, the two Neel walls shift along $+X$ and $-X$ creating
equal and opposite viscous forces in the $X$ direction.

The two peripheral Neel walls have the functional form
$\phi(x,y,X,Y) = f(x+y-X+Y\pm w)$, so that their contributions are
the same as that of the central Neel wall perpendicular to them.
Adding the contributions of all four Neel walls yields a total
\begin{equation}
\Gamma_{XX} = \Gamma_{YY} = - 2\Gamma_{XY} = 0.608 \alpha J t
w/\lambda, \label{eq-Gamma}
\end{equation}
independently of the vortex position.

It is instructive to compute the ratio of the viscous and gyrotropic
forces:
\begin{equation}
\Gamma_{XX}/G = 0.097 \alpha w/\lambda.
\label{eq-Gamma-over-G}
\end{equation}
The small value of Gilbert's damping in permalloy, $\alpha \approx 0.008$,
\cite{Freeman98} leads to the dominance of the gyrotropic force in strips
with submicron widths.  The smallness of $\Gamma_{XX}/G$ can be exploited
to organize an expansion in powers of this small parameter.

\section{Wall dynamics}
Equations of motion (\ref{eq-main}) for two generalized
coordinates $\xi_1 = X$ and $\xi_2=Y$ read
\begin{equation}\label{eom}
    F_i-\Gamma_{ij}\dot{\xi}_j + pG\epsilon_{ij}\dot{\xi}_j=0
\end{equation}
where $\Gamma_{ij}=\Gamma_{ji}$ is a viscosity tensor, $p$ is the
polarization of the vortex core, and $\epsilon_{ij}$ is the
$2 \times 2$ antisymmetric tensor with $\epsilon_{12}=+1$.\cite{Tretiakov07a}
The generalized forces $F_i=-\partial U/\partial \xi_i$ are derived
from the free energy (\ref{eq-U}).  We thus arrive at the following
equations of motion for the vortex core:
\begin{eqnarray}
    \dot{X}&=&\frac{Q H}{\Gamma_{XX}}+\frac{k(\Gamma_{XY}- p
    G)}{\det\Gamma+G^2}\left(Y-Y_{\mathrm{eq}}\right),\nonumber\\
    \dot{Y}&=&\frac{-k\Gamma_{XX}}{\det\Gamma+G^2}\left(Y-Y_{\mathrm{eq}}\right),
\end{eqnarray}
where the equilibrium $Y$ position of the vortex is given by
\begin{equation}\label{eq}
    k Y_\mathrm{eq}=- pG\frac{QH}{\Gamma_{XX}} (1+pg),
\end{equation}
where $g = (2\Gamma_{XX}+\Gamma_{XY})/G \ll 1$.  It is worth noting
that the magnitudes of the transverse displacement $|Y_\mathrm{eq}|$
are slightly different for the two values of the vortex polarization
$p$.  This effect can be traced to the lack of the reflection symmetry
$y \mapsto -y$ in a vortex wall, which leads to nonzero transverse
components of the Zeeman force $-2QH$ and the viscous force
$\Gamma_{YX}\dot{X}$.  As a result, trajectories of vortex cores with
$p = +1$ and $-1$ are slightly different.

Analysis of the equations of motion yields three distinct regimes
(Fig.~\ref{fig-mobility}).  Below a critical field $H_c$ we find
steady viscous motion with a high mobility $\mu = dV/dH$.  Immediately
above the critical field $H_c$ the motion exhibits an oscillatory
component; the drift velocity quickly decreases as the applied field
grows.  At much higher fields, $H \gg H_0$, the drift velocity
rises linearly again but with a much lower mobility $\mu$ than
at low fields.  The separation of scales $H_c$ and $H_0$ is
guaranteed by the smallness of the parameter $\Gamma_{XX}/G$.

\subsection{Low field: $H<H_c$}

In a low applied field the wall exhibits simple viscous motion.
The transverse coordinate of the vortex will asymptotically approach its
equilibrium position $Y_\mathrm{eq}$, so long as the latter is within
the strip.  The wall then moves rigidly with a steady longitudinal velocity
\begin{equation}
\dot{X} = QH/\Gamma_{XX}.
\label{eq-lfm}
\end{equation}
Experimental data of Beach \textit{et al.} \cite{Beach05} yield
\mbox{$Q/\Gamma_{XX} \sim 25$ (m/s) $\mathrm{Oe}^{-1}$} at low
fields for a strip 600 nm wide, which gives $\Gamma_{XX}/G = 0.13$.
while our Eq.~(\ref{eq-Gamma-over-G}) yields $\Gamma_{XX}/G = 0.12$
if we use the value of $\alpha=0.008$ measured by Freeman \textit{et al.}
\cite{Freeman98}

\subsection{Critical field: $H=H_c$}

The low-field regime ends when the equilibrium position of the vortex
core is pushed outside the strip edge, $|Y_\mathrm{eq}| \geq w/2$,
making the steady state unreachable.  As pointed out above, in permalloy
strips with a width below 1 $\mu$m the viscous force is
small in comparison with the gyrotropic one.  As a result, the
equilibrium of a vortex in the transverse direction is set mostly by
the balance of the transverse components of the gyrotropic force
$GV$ and the restoring force $-kY_\mathrm{eq}$.
The critical point is reached when $Y_\mathrm{eq} =w/2$:
\begin{equation}
GV_c = kw/2.
\label{eq-crit}
\end{equation}
With the aid of Eq.~(\ref{eq-k}) we obtain the critical velocity
\begin{equation}
V_c = \frac{1-1/\sqrt{5}}{2\pi^2} \gamma M t
\label{eq-vc}
\end{equation}
and the critical field
\begin{equation}
H_c = \Gamma_{XX}V_c/Q = kw \Gamma_{XX}/(2QG).
\label{eq-Hc}
\end{equation}

For permalloy, $\gamma = 2.21 \times 10^5 \mathrm{\ m \, A^{-1} \,
s^{-1}}$ and $M = 8.6 \times 10^5 \mathrm{\ m^{-1} \, A}$.
\cite{Stoll06} Taking the thickness of $t=20$ nm we obtain $V_c =
106$ m/s.  This is not too far from the critical velocity of 80 m/s
observed by Beach \textit{et al.}\cite{Beach05}

Equation (\ref{eq-vc}) shows that the critical velocity should grow
linearly with the film thickness $t$.  It is easy to see that this
result is valid beyond the crude model of a vortex wall adopted in
this calculation.  The two forces balancing each other
(\ref{eq-crit}) scale differently with $t$.  While the gyrotropic
force is linear in $t$, the restoring force comes from the
magnetostatic energy, which represents Coulomb-like interaction of
charges with density $\mathcal O(t)$, hence (the dipolar part of)
the restoring force is quadratic in $t$.  That gives $V_c \propto
t$.

\subsection{High field: $H>H_c$.  General remarks}

\begin{figure}
\includegraphics[width=0.9\columnwidth]{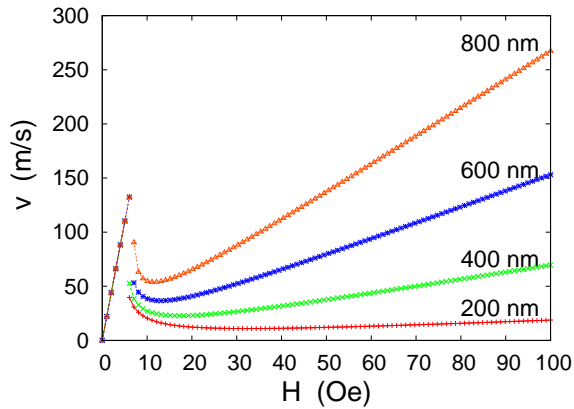}
\caption{Drift velocity vs. applied field curves for permalloy
strips of various widths. We use $t=20$ nm, $\alpha=0.01$, and $M =
8.6 \times 10^5 \mathrm{\ m^{-1} \, A}$.} \label{fig-mobility}
\end{figure}

Numerical simulations indicate that, after the original vortex with
a core polarization $p$ is expelled from the strip, a new vortex is
injected at the same location with the opposite polarization $-p$.
The vortex thus moves between the edges switching its core polarization
each time it reaches an edge.

Once the transverse coordinate of the vortex $Y$ becomes a dynamical
variable, the motion acquires an entirely different character.  As we
already pointed out, the gyrotropic force $\hat{G} \mathbf V$ dwarfs
the viscous one, $\hat{\Gamma} \mathbf V$, in permalloy strips.  To
zeroth order in $\Gamma_{XX}/G$, the dynamics is conservative: the
vortex core moves along equipotential lines $U(X,Y) = \mathrm{const}$.
At this order, the wall would oscillate back and forth but would not
move on average.  Drift requires a nonzero viscosity: as the wall
coordinate $X$ increases on average, the loss of Zeeman energy must
be accounted for through viscous friction.

\subsection{Very high field: $H \gg H_0$}

We first demonstrate that at a very high field the velocity is again
proportional to the field and calculate the high-field mobility.
The new field scale $H_0$ is set by the requirement that the restoring
force $-kY$ be negligible in comparison with the Zeeman force $QH$.
The characteristic field is
\begin{equation}
H_0 = kw/(2Q) = H_c G/\Gamma_{XX} \gg H_c.
\end{equation}
When $H \gg H_0$, the dynamics is dominated by the Zeeman and gyrotropic
forces, so that the vortex moves along an equipotential line
$Y=X/2 + \mathrm{const}$, or $\dot{X} = 2 \dot{Y}$.

As a result of the drift with a velocity $V_d$, the Zeeman energy
goes down on average at the rate $QH V_d$.  It is dissipated through
heat generated at the rate
\[
\mathbf V^T \hat{\Gamma} \mathbf V = \dot{Y}^2
\left(\begin{array}{cc}2 & 1\end{array}\right)
\hat{\Gamma}
\left(\begin{array}{c}2 \\ 1\end{array}\right).
\]
The transverse velocity of the vortex core reflects the balance between
the longitudinal components of the gyrotropic and Zeeman forces:
$\dot{Y} \approx QH/G$.  We thus find the drift velocity
\begin{equation}
    V_d = \frac{QH}{G^2}\left(\Gamma_{YY}
    +4\Gamma_{XX}+4\Gamma_{XY}\right) = \frac{3\Gamma_{XX}QH}{G^2}.
    \label{eq-hfm}
\end{equation}
In the last transition we have used the relation between the coefficients
of the viscosity tensor specific to this model (\ref{eq-Gamma}).

The high-field (HF) mobility (\ref{eq-hfm}) is suppressed in comparison
to the low-field (LF) one (\ref{eq-lfm}):
\begin{equation}
\frac{\mu_\mathrm{HF}}{\mu_\mathrm{LF}} = \frac{3 \Gamma_{XX}^2}{G^2}
\ll 1.
\end{equation}
In the experiment of Beach \textit{et al.},\cite{Beach05}
$\mu_\mathrm{HF}/\mu_\mathrm{LF}\approx 0.1$, while the theoretical
result is $3(\Gamma_{XX}/G)^2\approx 0.05$, \textit{i.e.} twice as
small.

\subsection{High field: $H>H_c$.  Details}
\label{full mobility} To find the drift velocity of the vortex at
fields above the vortex expulsion field $H_{c}$, we determine the
total $X$ displacement of the vortex over a full cycle of motion
from the top of the strip to the bottom and back again. The crossing
time will be slightly different on the upward and downward trips due
to the asymmetry of the vortex wall and the $Y$ component of the
Zeeman force.

Solving Eq.~(\ref{eom}) with polarization $p=\pm 1$ gives us the
crossing times and displacements $\Delta T_+$ and $\Delta X_+$ (top to
bottom) and $\Delta T_-$ and $\Delta X_-$ (bottom to top):
\begin{eqnarray}
    \Delta X_\pm&=& \frac{QH \Delta T_\pm}{\Gamma_{XX}}
    -\frac{G \mp \Gamma_{XY}}{\Gamma_{XX}}w,
\nonumber\\
    \Delta T_\pm&=&\frac{\det\Gamma+G^2}{k\Gamma_{XX}}
    \ln\left(\frac{1+H_c/H \pm g}{1-H_c/H \pm g}
    \right).
\end{eqnarray}
The drift velocity is
\begin{eqnarray}
    V_d &=& \frac{\Delta X_+ +\Delta X_-}{\Delta T_+ + \Delta T_-}
\label{eq-vd}
\\
    &=&V_c\left(
      \frac{H}{H_c}
      - \frac{4}{(1+\det{\Gamma}/G^2)
          \ln\left(\frac{(1+H_c/H)^2-g^2}{(1-H_c/H)^2-g^2}\right)}
    \right).
\nonumber
\end{eqnarray}
The resulting curve is shown for several strip widths in
Fig.~\ref{fig-mobility}. Note that the critical field is not exactly
$H_c$ and actually changes slightly with the width. This is because
the equilibrium points for both up- and downwardly polarized
vortices must be expelled from the strip for the character of the
motion to change. By Eq.~(\ref{eq}) a downwardly polarized vortex
requires a slightly higher field to expel than an upwardly polarized
one. An expansion of Eq.~(\ref{eq-vd}) in powers of $1/H$ yields the
high-field result (\ref{eq-hfm}).

\section{Discussion}

We have explored the dynamics of a vortex domain wall in a magnetic
strip of a submicron width.  We have applied the method of
collective coordinates\cite{Tretiakov07a} to the case when
the wall has two soft modes related to the motion of the vortex
core.  A simplified model of the vortex domain wall described in
this paper yields solvable equations of motion.  The calculated mobility
of the wall in the steady-state viscous regime at low fields agrees well
with the value measured by Beach \textit{et al.}\cite{Beach05}
The steady motion breaks down when the equilibrium position of the vortex
moves beyond the edge of the strip.  The critical velocity
(\ref{eq-vc}) depends just on the magnetization length and the
sample thickness; its calculated value agrees reasonably well with
the data of Beach \textit{et al.}\cite{Beach05}  The dynamics above
the breakdown changes the character from overdamped to underdamped:
the ratio of the viscous and gyrotropic forces acting on the wall
$\Gamma_{XX}/G = 0.13$ in their experiment.  In this regime the
velocity sharply declines at first but later starts to rise again as
the field strength increases. The high-field mobility is reduced in
comparison with the low-field value by the factor $3\Gamma_{XX}^2/G^2
=0.05$; the observed reduction is not as strong:
$\mu_\mathrm{HF}/\mu_\mathrm{LF} \approx 0.1$.\cite{Beach05}

In addition to simplifying the geometry (but not the topology) of
the domain wall, we have made other assumptions that require further
checking. First, we have assumed that any vortex absorbed by the
edge is immediately reemitted. At fields lower than that required
for emission to occur, the wall may simply stay transverse and
continue to move in a viscous fashion. At higher fields, there may
be short delays between absorption and reemission during which the
motion of the wall is again viscous; the higher mobility of a
transverse wall would tend to increase the drift velocity.

Second, just as the appearance of $Y$ as a new degree of freedom
completely changes the character of the wall dynamics above the
critical field $H_c$, at still higher fields additional modes of the
wall may become important. The number and dynamical characteristics
of soft modes may also change discontinuously as additional vortices
or antivortices are created and annihilated in the bulk of the
strip. We have observed the creation and subsequent annihilation of
a vortex-antivortex pair near the original vortex of the wall.  Like
the process described by Van Waeyenberge \textit{et
al.,}\cite{Stoll06} the pair creation mediates the flipping of the
polarization of the wall vortex and results in the reversal of the
gyrotropic force.  Thus the dynamics is similar to that described in
this paper: the vortex moves back and forth, while the domain wall
slowly drifts along the strip. A possible way to detect this new
regime is to measure the frequency of longitudinal oscillations:
because the vortex does not reach the edge, the frequency should be
\textit{higher} than $2\omega =2\gamma H$ expected when the vortex
moves from edge to edge.\cite{Tretiakov07a} We shall describe the
onset of this type of motion more fully in future work.
\section{Acknowledgments}

We thank Ya. B. Bazaliy, G. S. D. Beach, and C. L. Chien for
valuable discussions.  This work was supported in part by the NSF Grant
No. DMR-0520491.

\bibliography{micromagnetics}

\end{document}